\documentclass[12pt]{article}
\usepackage{epsfig}

\newcommand{\tdm}[1]{\mbox{\boldmath $#1$}}

\newcommand{\f}[2]{\frac{#1}{#2}}

\newcommand{\eq}{\begin{equation}}
\newcommand{\eqx}{\end{equation}}
\newcommand{\eqn}{\begin{eqnarray}}
\newcommand{\eqnx}{\end{eqnarray}}

\newcommand{\lab}{\label}

\newcommand{\al}{\alpha}

\title{{\bf IMPLEMENTATION OF THE RECOVERING CORRECTIONS INTO
            THE INTERMITTENT DATA ANALYSIS}
       }
\author{Beata Ziaja \footnote{e-mail: {\tt beataz@qcd.ifj.edu.pl}}\\
        \it Department of Theoretical Physics\\
        \it Institute of Nuclear Physics\\
        \it Radzikowskiego 152, 31-342 Cracow, Poland\\}
\date{March 1999}

\begin{document}

%
%
%
%

\maketitle

\begin{abstract}
The improved method of intermittent data analysis is proposed.
It exploits, in addition to the standard density moments, the
information on the bin-bin correlations, observed in the data
and expressed in terms of the density correlators.
The improving recovering corrections are implemented into the 
data analysis in the form of the recursive algorithm, and 
tested in the framework of multiplicative cascading models.
\end{abstract}
The first signals on possible intermittent behaviour in high-energy 
multiparticle production \cite{l1} were found in the data of the single event 
recorded by the JACEE collaboration \cite{l2}. The presence
of large dynamical fluctuations manifesting a scaling behaviour was
registered also afterwards in other accelerator experiments \cite{acc}.
Many different models \cite{l4} have been proposed since to explain the 
power-law rise of the multiparticle moments, described by the coefficients
called the intermittency exponents. Some of the models suggested that 
the observed scaling may be the result of  final state multiparticle 
cascading \cite{l4}, and the intermittent data represent the last stage of 
the cascade. In this approach the main problem lies in the 
extraction of the information on the previous cascade stages 
which are in some way encoded in the last stage data. 
The standard method of recovering the history of the cascade 
was proposed and applied originally to the JACEE 
event data. Since that time it has become a standard tool of multiparticle data 
analysis \cite{l4}, especially in the event-by-event analysis \cite{chin,bz}.
However, the theoretical problem what is the interplay between 
the cascade recovered in the standard analysis and the true cascade 
which generated the data has been studied only recently \cite{jz}, 
\cite{ziaja}. It ended up with the proposal of introducing the new 
recovering corrections to the data analysis.

In this paper we would like to summarize the  results on improvement of the standard data
analysis achieved by including the recovering corrections. 
Most of these results were derived in Ref.\ \cite{ziaja}. In what follows we will concentrate 
on the technique of
implementing the recovering corrections into the data analysis in the way 
which may be useful for experiment.

The {\bf recovering corrections} aim to improve the method of
recovering the history of the particle cascade, also called the
rebinning \cite{l4}.
The standard method reconstructs the cascade from the last step data,
represented by a sample of M numbers~: $x^{(n)}_i$ ($i=1,\ldots,M$). They
describe e.\ g.\ the distribution of particle density into bins. For simplicity
assume $M=2^n$, where $n$ denotes the number of cascade steps. The recovered
particle density $y^{(n-k)}_i$  in the $i$th bin at the
$(n-k)$th cascade step ($k=0,\ldots,n-1$) takes then the form~:
\eq
y^{(n-k)}_i= \f{1}{2^k} \sum_{j=0}^{2^k-1}
x^{(n)}_{2^k\times i+j}.
\lab{r5}
\eqx
The intermittency exponents are extracted from unnormalized
{\bf reconstructed density moments} $z_{q;\,rec.}^{(k)}$
\footnote{Actually, the factorial moments are normally used (to reduce the
statistical noise) but this does not concern us here}~:
\eq
z^{(k)}_{q;\,rec.}=\f{1}{2^k} \sum_{j=0}^{2^k-1}  \left(
y^{(k)}_j \right)^q,
\lab{r6}
\eqx
assuming that $z_{q;\,rec.}^{(k)}$ manifests a power law behaviour~:
\eq
z^{(k)}_{q;\,rec.} \sim 2^{k\cdot \phi'_q},
\lab{r7}
\eqx
and the normalized intermittency exponent $\phi_{q;\,norm.\ }=\phi_q'$,
where $\phi_{q;\,norm.\ }:=\phi_q-q\phi_1$. The exponent 
$\phi_q'$ is estimated as a slope of the linear $\chi^2-$fit 
applied to the points $(k,\log z^{(k)}_{q;\,rec.})$ ($\log x \equiv \log_2 x$).

However, it was found and proved in \cite{jz}, \cite{ziaja} that there
exists a difference between the true density moments~:
\eq
z^{(k)}_q=\f{1}{2^k} \sum_{i=0}^{2^k-1} \left( x^{(k)}_i \right)^q
\lab{r2}
\eqx
and the reconstructed ones (\ref{r7}) which influences the estimation
of intermittency exponents. This difference may be expressed in the
form of correcting factor $p_q(k)$~:
\eq
z^{(n-k)}_{q;\,rec.}=
z^{(n-k)}_q \cdot p_q(k),
\lab{r8}
\eqx
which was called the {\bf recovering correction} \cite{ziaja}.

The recovering corrections contain information on the specific process
which generated the true cascade. In what follows we restrict ourselves
to the corrections considered in the framework of multiplicative
random cascading models \cite{models,bsz,MS87} which are nowadays widely recognized in
multiparticle data analysis \cite{l4}.
Similarly as in \cite{ziaja}, we consider the multiplicative
models with possible neighbour-node memory which generate a uniform
distribution of particle density into bins. In the multiplicative cascade
the root of the cascade is set equal 1~: $x_0^{(0)}=1$.
One generates the next stages of the cascade recursively,
following the scheme~:
\eqn
x^{(k+1)}_{2i}   &:=& W_1 \cdot x^{(k)}_i,\nonumber\\
x^{(k+1)}_{2i+1} &:=& W_2 \cdot x^{(k)}_i ,
\lab{r14}
\eqnx
where $W_1$ and $W_2$ are random variables of  $m$ model parameters
$a_j$, $j=1,\ldots,m$~:
\begin{center}
$W_1=a_j$ with probability $p_{a_j}$,\\
$W_2=a_j$ with probability $p_{a_j}$,
\end{center}
\eq
\lab{r1414}
\eqx
with normalized probability weights $\sum_{j=1}^{m}p_{a_j}=1$.
The distribution of particle density will be uniform if the following
condition is fulfilled~:
\eq
p(W_1=a_i,W_2=a_j)=p(W_1=a_j,W_2=a_i),
\lab{unif}
\eqx
where $p(W_1=a_i,W_2=a_j)$ denotes probability of choosing in (\ref{r14})
$W_1=a_i$ and $W_2=a_j$ ($i,j=1,\ldots,m$).
Then unnormalized density moments $z_q^{(k)}$ fulfill the scaling relation~:
\eq
z^{(k)}_q \sim 2^{k\cdot \phi_q},
\label{r3}
\eqx
and intermittency exponents $\phi_q$ read~:
\eq
\phi_q=\log(a_1^q p_{a_1}+\ldots+a_m^q p_{a_m}).
\lab{r17}
\eqx
The popular models~: $\al-$, $p-$models \cite{models},\cite{MS87} and the
$(p+\al)-$model introduced in \cite{ziaja} are special cases of
multiplicative rule (\ref{r14}).

It was proved in \cite{ziaja} that for any multiplicative process
which obeys rules (\ref{r14}), (\ref{unif}) 
recovering corrections $p_q(k)$ fulfill the recurrence equation
\footnote{A similar recurrence relation has been obtained in a different context
in \cite{pesch}}~:
\eq
p_q(k)=\f{1}{2^{q}}\sum_{j=0}^q \left(\begin{array}{c} q\\j\end{array}
\right) p_j(k-1) p_{q-j}(k-1) \langle W_1^jW_2^{q-j}\rangle
\lab{r19}
\eqx
with the initial conditions~:
\eqn
p_q(0)&=&1,\nonumber\\
p_0(k)&=&1.
\lab{r1919}
\eqnx
We introduce a notation~:
\eq
\langle W_1^jW_2^{l}\rangle \equiv k_{j,l}.
\lab{not}
\eqx
Formula (\ref{r19}) implies that coefficients $k_{j,l}$ are the only parameters
of the multiplicative model needed for calculating the value of $p_q(k)$.
Furthermore, it is not difficult to
establish the values of $k_{j,l}$ from the model. 
For either $j=0$ or $l=0$ they equal~:
\eq
k_{j,0}=k_{0,j}=2^{\phi_j},
\lab{k0}
\eqx
where $\phi_j$'s are ordinary intermittency exponents (\ref{r3}).
To find the value of $k_{j,l}$ for both $j,l\neq 0$
we use the {\bf unnormalized density correlators}
$c_{j,l}^{(k)}$ \cite{l1,l4,corr}~:
\eq
c^{(k)}_{j,l}=\f{1}{2^{k-1}}
\sum_{i=0}^{2^{k-1}-1}
\left( x^{(k)}_{2i} \right)^j \left( x^{(k)}_{2i+1} \right)^l.
\lab{defcor}
\eqx
The correlators and the density moments fulfill the relation (see
\cite{ziaja})~:
\eq
c^{(k)}_{j,l}=z^{(k-1)}_{j+l}\cdot k_{j,l},
\lab{cz}
\eqx
which can be also rewritten as~:
\eq
\log c^{(k)}_{j,l}=(k-1)\phi_{j+l}+\log k_{j,l}.
\lab{cz2}
\eqx
Both relations (\ref{cz}), (\ref{cz2}) imply that we may derive $k_{j,l}$
in a straigthforward way by calculating correlators and density moments
from data, and applying to them the standard $\chi^2-$fit.

Applying the standard method to the correlators at the previous cascade stages,
we would expect to find the similar difference between the
reconstructed correlators and the true ones, as it was observed for the
density moments. It was proved in \cite{ziaja} that this difference
may be expressed in terms of the same recovering correction $p_q(k)$ 
(\ref{r19}) as for the density moments~:
\eq
c^{(n-k)}_{j,l;\,rec.}=c^{(n-k)}_{j,l}p_{j+l}(k).
\lab{crec}
\eqx

Now we have all tools needed for implementation of recovering corrections
into the multiplicative data analysis. To illustrate the problem
we describe the improved estimation of intermittency exponents
of the second rank. The recovering correction of the second rank derived
from (\ref{r19}) reads~:
{\footnotesize
\eq
p_2(k)={1 \over 4}(\,p_2(k-1)k_{2,0}+p_2(k-1)k_{0,2}+2p_1^2(k-1)k_{1,1}\,)
\lab{p2}
\eqx}
with the initial condition $p_2(0)=1$. It was proved in \cite{ziaja} that~:
\eq
p_1(k)=2^{k\,\phi_1}=\left( z_1^{(n)}\right)^{k\over n}
\lab{p1}
\eqx
The parameters needed to calculate $p_2(k)$ are following~: $z_1^{(n)}$,
$\phi_2$ (cf.(\ref{k0})) and $k_{1,1}$. Derivation of $z_1^{(n)}$
is straigthforward, and to estimate the values of
$\phi_2$ and $k_{1,1}$ we propose the following recursive procedure.
The primary values of $\phi_2$ and $k_{1,1}$ may be obtained in the standard
way from (\ref{r7}),(\ref{cz}). We  substitute them to formula (\ref{p2})
to derive the approximate form of correction $p_2(k)$. Now, applying
again the approximate form of $p_2(k)$ to equations (\ref{r7}),
(\ref{cz}), one derives adjusted parameters $\phi_2$ and $k_{1,1}$
and compares them with the primary values. If the relative difference
is large, one repeats the recursive adjusting
till the parameters do not change within a given accuracy.

One may generalize the above scheme for the intermittency exponents
of any rank. Following \cite{ziaja}, below we present the implementation
algorithm which recursively adjusts the primary parameters
$\phi_q$, $k_{j,l}$ ($j+l=q$, $jl>0$) obtained after applying the standard
method to the data~:\\

({\bf INPUT}) parameters $\tdm\phi_1,\ldots,\tdm\phi_{q-1}$,
$\tdm k_{j,l}$ ($j+l=1,\ldots,q-1$)
obtained after applying the implementation algorithm for
$q=1,2,\ldots,q-1$ step-by-step\\

({\bf 1}) derive $\tdm\phi'_q$, $\tdm k'_{j,q-j}$ ($j=1,\ldots,q-1$)
from data, using the standard method\\ i.\ e.\ reconstruct the cascade
using (\ref{r5}) and derive the parameters from relations~:
\eq
\label{al1}
\log z^{(k)}_{q;\,rec.\ } = k\cdot \phi'_q + b.
\eqx
\eq
c^{(k)}_{j,l;\,rec.\ }=z^{(k-1)}_{j+l;\,rec.\ }\cdot k'_{j,l}
\lab{al2}
\eqx
where $k=1,\ldots,n$ (cf.\ (\ref{r7}), (\ref{cz})),\\

({\bf 2}) derive $\tdm\phi_{q;\,corr.\ }$, $\tdm k_{j,q-j;\,corr.\ }$
($j=1,\ldots,q-1$) in the following substeps~:\\

\noindent
({\small \bf 2.0}) calculate $p_q(k)$ from relation (cf.\ (\ref{r19}))~:
\eqn
p_q(k)=\f{1}{2^{q}}\sum_{j=0}^q \left(\begin{array}{c} q\\j\end{array}
\right) p_j(k-1) p_{q-j}(k-1) k_{j,q-j},
\lab{al3}
\eqnx
using $\tdm\phi'_q$, $\tdm k'_{j,q-j}$ derived in step
({\bf 1}), and estimate $\tdm\phi_{q;\,corr.}$ from~:
\eqn
\label{al4}
\log z^{(n-k)}_{q;\,rec.} -\log(p_q(k)) = (n-k)\cdot \phi_{q;\,corr.} +b,
\eqnx\\

\noindent
({\small \bf 2.1} ) calculate $p_q(k)$ from (\ref{al3}) using $\phi_{q;\,corr.}$ (other parameters
as after step (1)), and estimate $\tdm k_{1,q-1;\,corr.}$ from relation
(cf.\ (\ref{cz2}),(\ref{al3}))~:
\eqn
\log c^{(n-k)}_{j,l;\,rec.}-\log(p_{j+l}(k))=(n-k-1)\phi_{j+l}+\log k_{j,l;\,corr.},
\lab{al5}
\eqnx

\noindent
$\ldots$,\\

\noindent
({\small \bf 2.q-1}) calculate $p_q(k)$ from (\ref{al3}), using all previously derived parameters
$\tdm\phi_{q;\,corr.\ }$, $\tdm k_{j,q-j;\,corr.\ }$, and
estimate $\tdm k_{q-1,1;\,corr.}$ from (\ref{al5}),\\

({\bf 3}) compare the values of $\tdm\phi'_q$, $\tdm k'_{j,q-j}$ and
    $\tdm\phi_{q;\,corr.}$, $\tdm k_{j,q-j;\,corr.}$ ($j=1,\ldots,q-1$). If the relative
    difference is large, assume~:
\eqn
\phi'_q   &:=&\phi_{q;\,corr.},\nonumber\\
k'_{j,q-j}&:=&k_{j,q-j;\,corr.}\nonumber
\eqnx
and repeat steps (2),(3) recursively until the relative difference
between parameters before and after step (2) is small enough. Then
go to the output, assuming $\tdm\phi_q:=\tdm\phi'_q$,
$\tdm k_{j,q-j}:=\tdm k'_{j,q-j}$   \\

({\bf OUTPUT}) parameters $\tdm\phi_1,\ldots,\tdm\phi_{q}$;
$\tdm k_{j,l}$ ($j+l=1,\ldots,q$).\\

Technical details and problems which appear when applying
the algorithm to data were discussed in detail in \cite{ziaja}.\\

We have performed numerical simulations of the $\alpha-$, $p-$ and
$(p+\al)-$models \cite{ziaja} in order to test how the implementation
algorithm works in practice.
We generated 10000 cascades of the 10 step length for the $\alpha-$
and $(p+\alpha)-$models, and one cascade
of the 10 step length for the $p-$model \footnote{It can be proved that for a given
parameter set the $p-$model generates always the same values of the correlators
and density moments} for two different parameter sets separately.

Implementation algorithm analized the data of the last cascade step.
For each event it estimated the value of normalized intermittency exponents
$\phi_{2;\,norm.}$, $\phi_{3;\,norm.}$ ($\phi_{i;\,norm.}:=\phi_i-i\cdot\phi_1$),
using the standard  method (step 1) with recovering corrections included
(steps 2,3).
The selected results (for one set of parameters) are presented in Fig.\ 1
and in Tabs.\ 1, 2.

For the $\al-$model the histograms of $\phi_{2;\,norm.}$, $\phi_{3;\,norm.}$
obtained in the standard method and the histograms with recovering corrections
included are almost identical (see Figs.\  1a, 1b and Tab.\ 1). In this case the recovering corrections can
be implemented better when one applies directly dedicated $\alpha-$model recovering
correction \cite{ziaja} i.\ e.\ if one substitutes coefficient $k_{j,q-j}$ in (\ref{r19})
by the product~: $k_{j,q-j}=2^{\phi_j}\cdot 2^{\phi_{q-j}}$.

On the contrary, the implementation algorithm works well for
the $(p+\alpha)-$model (see Figs.\ 1c, 1d and Tab.\ 2).
For the $(p+\alpha)-$model the histogram with the recovering corrections
included approximates well the theoretical value of normalized
intermittency exponent. The histogram obtained by using the standard
method is moved slightly to the left in comparison to the histogram with
recovering corrections included.

We have checked that for the $p-$model the theoretical values of normalized
intermittency exponents are estimated perfectly by both standard method
and implementation algorithm \cite{ziaja}.

It should be also mentioned that the histograms generated by the
implementation algorithm (recovering corrections) are symmetric, in
contrast to the standard ones, and their dispersions are of the same order
as those derived for the standard method (cf. Tabs.\ 1, 2).

To sum up we analyzed the estimation of intermittency exponents from the data
which were generated by a multiplicative random cascading process.
The following methods were applied: the standard method of cascade
recovering (\ref{r5}) and the improved method which included
recursively the recovering corrections. The improved method was applied
in the form of the implementation algorithm. Numerical simulations have been
performed to check how both methods work in practice.
The conclusions may be summarized as follows~:\\

(a)
standard method of estimation of intermittency exponents does not apply
for the whole class of multiplicative models~: its accuracy
depends on the specific properties of the model
and its parameters. The method does not detect a conservation law
if present in the model;\\

(b)
we propose an improved method of estimation of intermittency exponents.
It exploits, in addition to the standard density (factorial) moments,
the information on the bin-bin correlations, observed in the data and
expressed in terms of the density correlators;\\

(c)
the method is formulated in the form of the recursive algorithm which, 
starting from the parameters obtained from the density moments and 
correlators, allows successive improvements of the result;\\

(d)
the method was tested in MC simulations which show that it is workable
and indeed brings the experimental estimates closer to their theoretical
values. Moreover, the improved distributions are symmetric with
approximately the same dispersions as the uncorrected ones.\\

\section*{Acknowledgements}

I would like to thank Prof.\ A. Bia{\l}as for reading the manuscript and
many suggestions and comments and Dr.\ R.\ Janik for discussions. 
This work was supported in part by Polish Government grant Project 
(KBN) 2P03B04214.

\newpage
\noindent
{\bf Figure captions}\\ \\
\noindent
{\bf Figs.\ 1a, 1b}
Estimation of normalized intermittency exponents $\phi_{2;\,norm.}$
and $\phi_{3;\,norm.}$
for $\al-$model, using the standard  method (dotted line), the improved method
with the implementation algorithm (thin solid line), and dedicated
$\al$-~corrections \cite{ziaja} (dashed line) compared with the theoretical
values (solid line), performed for one set of $\al-$model parameters~:\\ \\
$a_1=0.8$, $a_2=1.1$, $p_1=1/3$\\  \\
{\bf Figs.\ 1c, 1d}
Estimation of normalized intermittency exponents $\phi_{2;\,norm.}$
and $\phi_{3;\,norm.}$
for $(p+\al)-$model, using the standard  method (dotted line),
the improved method with the implementation algorithm (thin solid line),
compared with the theoretical values (solid line), performed for
one set of $(p+\al)-$model parameters~:\\ \\
      $a_{2i}=1-a_{2i-1}$, $p_{2i}=p_{2i-1}$ for $i=1,\ldots,10$,\\
      $a_1   =0.2$, $a_3   =0.5$, $a_5   =0.6$, $a_7   =0.3$, $a_9   =0.45$,\\
      $a_{11}=0.25$, $a_{13}=0.1 $, $a_{15}=0.15$, $a_{17}=0.87$, $a_{19}=0.66$,\\ \\
      $2 p_1=0.05$, $2 p_3=0.15$, $2 p_5=0.25$, $2 p_7=0.40$, $2 p_9=0.05$,\\
      $2 p_{11}=0.05$, $2 p_{13}=0.02$, $2 p_{15}=0.02$, $2 p_{17}=0.005$, $2 p_{19}=0.005$,\\ \\
\newpage
\begin{table}[hbpt]
\noindent
{\bf Tab.\ 1} Estimation of normalized intermittency exponents
$\phi_{2;\,norm.}$ and $\phi_{3;\,norm.}$ and their dispersions
for the $\alpha-$model, using the standard  method (second column),
the improved method with the implementation algorithm (third column),
and dedicated  $\alpha-$~corrections \cite{ziaja} (fourth column),
compared with the theoretical values (first column), performed for
one set of $\alpha-$model parameters (cf.\ Figs.\ 1a, 1b)~:

\begin{center}
\begin{tabular}{|r|c|c|c|c|c|c|}
\hline \hline
                & theor.&standard&algorithm& $\alpha$-corr.\\
\hline
   $\phi_{2;\,norm.}$   &$0.0285$&$0.0251\pm0.004$&$0.0246\pm0.0033$&$0.0288\pm0.004$\\
\hline
   $\phi_{3;\,norm.}$   &$0.0813$&$0.0757\pm0.010$&$0.0727\pm0.009$&$0.0798\pm0.0111$\\
\hline
\hline
\end{tabular}
\end{center}
\end{table}
\begin{table}[hbpt]
\noindent
{\bf Tab.\ 2} Estimation of normalized intermittency exponents
$\phi_{2;\,norm.}$ and $\phi_{3;\,norm.}$ and their dispersions
for the $(p+\alpha)-$model, using the standard  method (second column),
the improved method with the implementation algorithm (third column),
compared with the theoretical values (first column), performed for
one set of $(p+\alpha)-$model parameters (cf.\ Figs.\ 1c, 1d)~:

\begin{center}
\begin{tabular}{|r|c|c|c|c|c|c|}
\hline \hline
                & theor.&standard&algorithm\\
\hline
   $\phi_{2;\,norm.}$   &$0.177$&$0.170\pm0.023$ &$0.173\pm0.029$\\
\hline
   $\phi_{3;\,norm.}$   &$0.478$&$0.438\pm0.069$ &$0.470\pm0.092$\\
\hline \hline
\end{tabular}
\end{center}
\end{table}
\newpage
\begin{figure}[t]
\hfill a)\epsfig{width=7.0cm, file=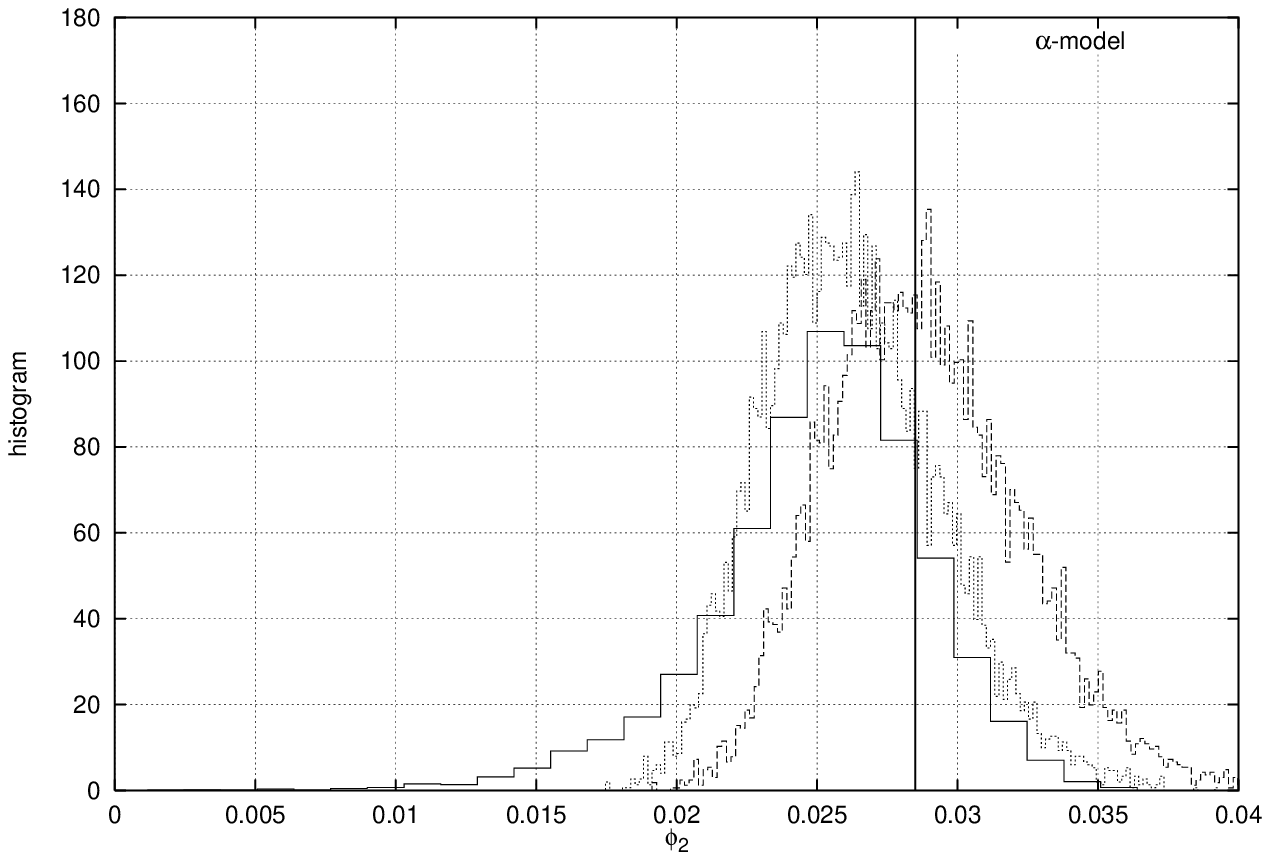}\hfill
b)\epsfig{width=7.0cm, file=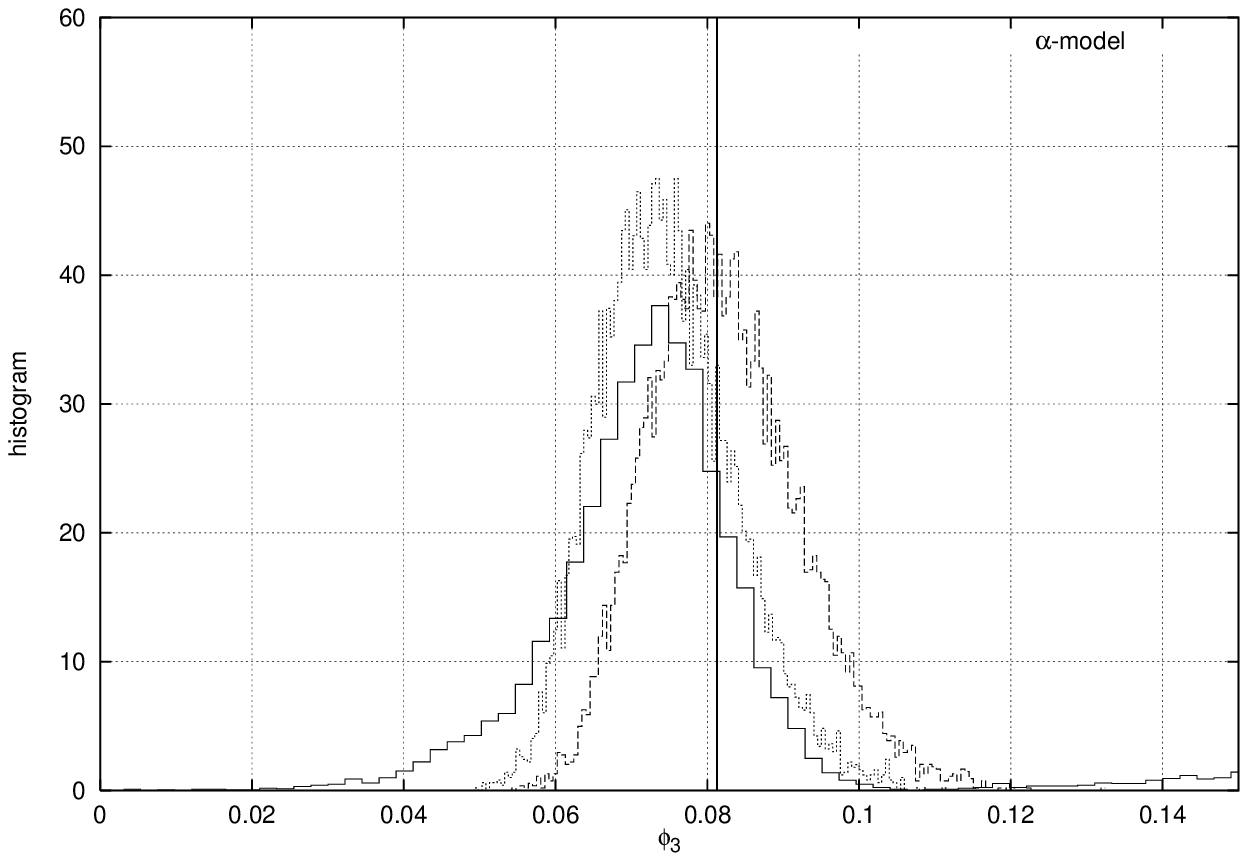}\hfill\mbox{}\\
c)\hfill\epsfig{width=7.0cm, file=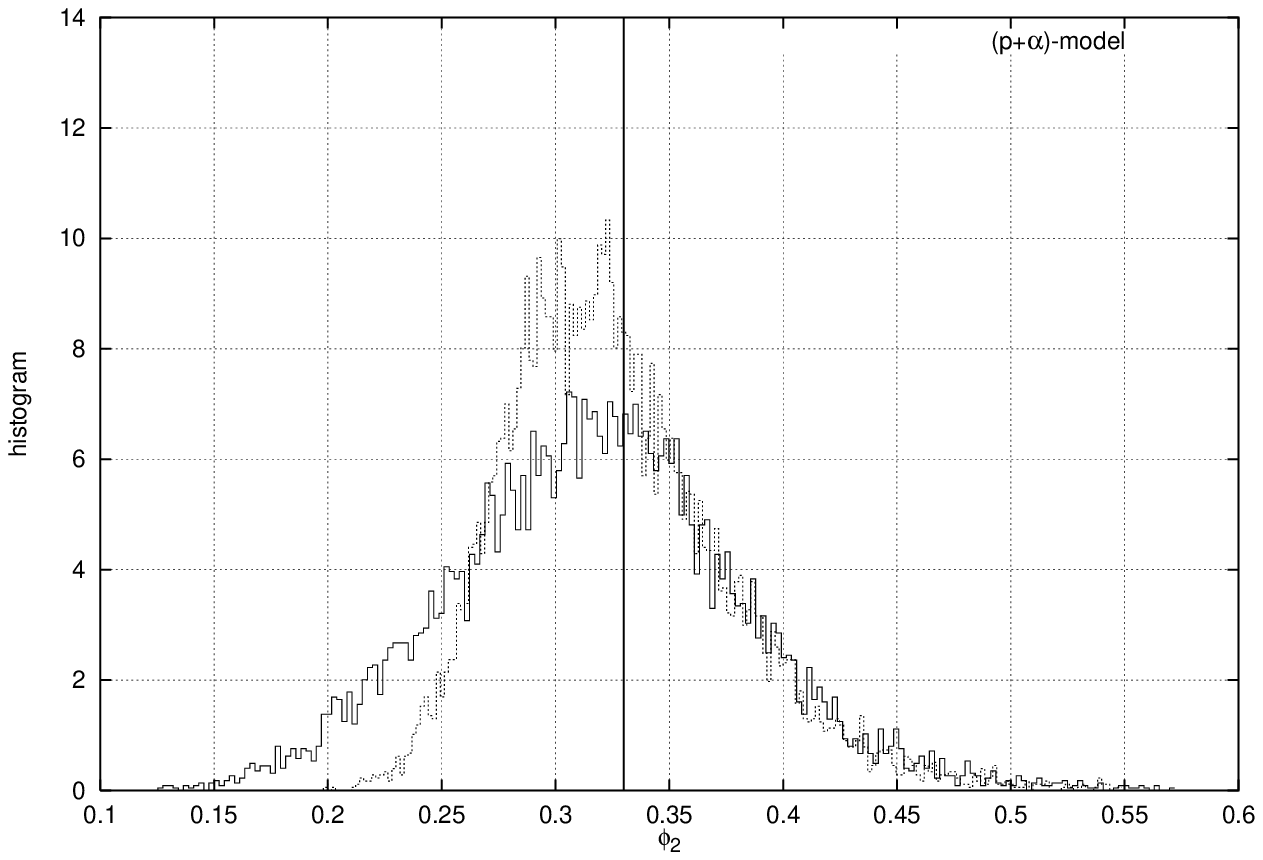}\hfill
d)\epsfig{width=7.0cm, file=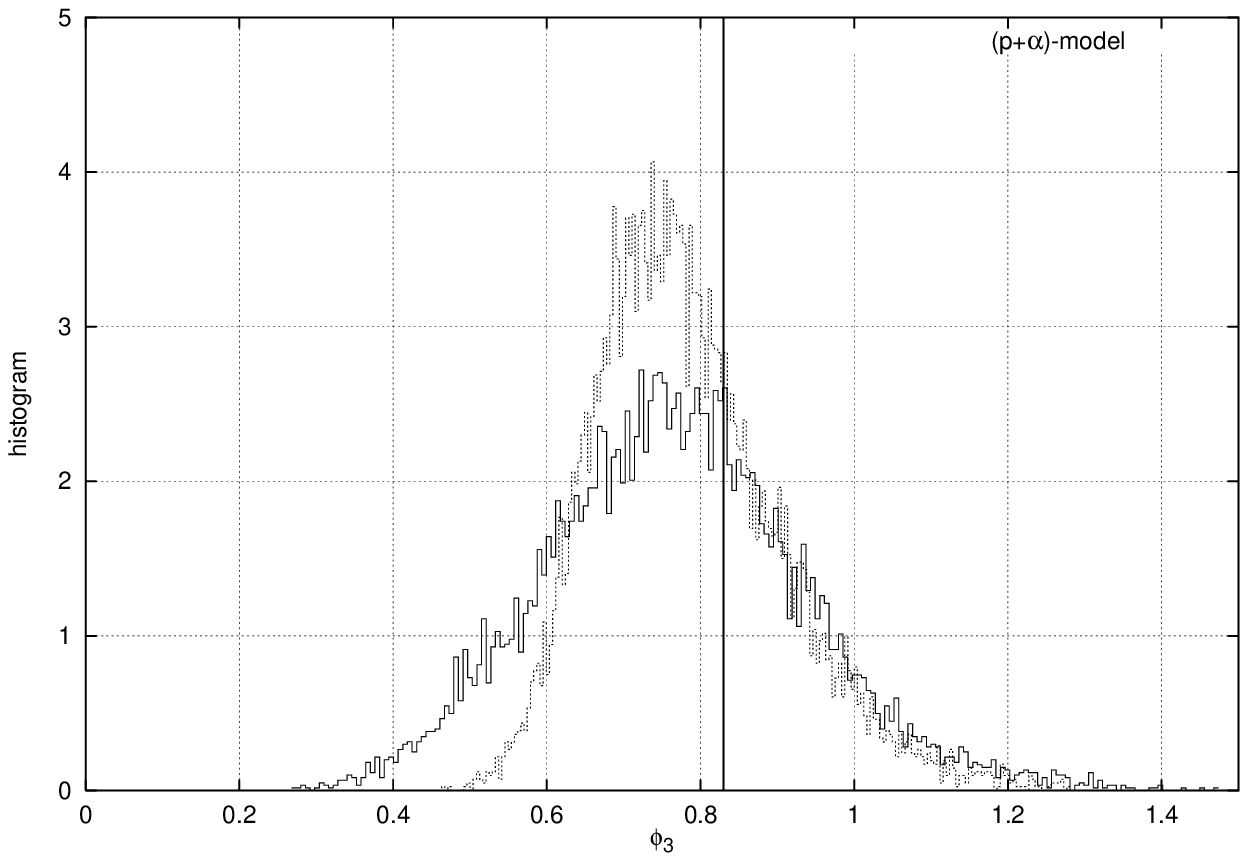}\hfill\mbox{}
\caption{}
\end{figure}
%
%
%
%
\end{document}